# Monitoring Wandering Behavior of Persons Suffering from Dementia Using BLE Based Localization System

Authors version

Marcin Kolakowski [1], Bartosz Blachucki[2]

[1]Institute of Radioelectronics and Multimedia Technology, Warsaw University of Technology, Warsaw, Poland, contact: marcin.kolakowski@pw.edu.pl

[2]Department of Geriatrics Warsaw, Medical University of Warsaw, Warsaw, Poland

# Monitoring Wandering Behavior of Persons Suffering from Dementia Using BLE Based Localization System

Marcin Kolakowski, *Student Member, IEEE,* Bartosz Blachucki, *MD*

*Abstract* — With the aging of our populations, dementia will become a problem which would directly or indirectly affect a large number of people. One of the most dangerous dementia symptoms is wandering. It consists in aimless walking and spatial disorientation, which might lead to various unpleasant situations like falling down accidents at home to leaving the living place and going missing. Therefore, in order to ensure elderly people's safety it is crucial to detect and alarm the caregivers in case of such incidents. It can be done by tracking the sufferers movements and detecting signs of repetitiveness. The paper presents the results of the study, in which the wandering behavior of people suffering from dementia was monitored using a Bluetooth Low Energy based positioning system. The paper includes the description of the system used for patients localization and the results of the tests performed in a long term care facility.

## I. Introduction

DEMENTIA is a set of different symptoms occurring due to brain disease (predominantly Alzheimer Disease) causing its progressive disorganization. These symptoms (e.g. memory loss, disturbed thinking and perception, inability to solve problems, language difficulties) affect daily life. Most people with dementia experience common behaviors, such as agitation, aggression, resistance to care, vocalizations, and wandering behavior [1]. The prevalence rate of dementia in Europe standardized for age and sex is 7.1%, based on the meta-analysis studies that were carried out between 1993 and 2018 [2].

Wandering behaviors of patients with dementia include aimless persistent walking, eloping behaviors and spatial disorientation. Ali et al. investigated adverse ramifications of this state in 143 dyads of persons with mild dementia and their caregivers from a veteran's hospital and memory clinic in Florida: a total of 49% of the study participants had falls, fractures, and injuries due to wandering behavior, and 43.7% demonstrated eloping behaviors [3].

Detecting of wandering behavior is crucial for caretakers of patients with memory loss. People suffering from dementia can easily get disoriented and lost. MacAndrew et al. searched news articles published in Australia between 2011 and 2015 reporting 130 missing persons diagnosed with dementia, 71% of the individuals were found. Of these, 60% were in good state, 20% were injured, and 20% were deceased [4]. This situation motivates the research teams all over the world to conduct research on wandering behavior and develop technological solutions to better cope with this problem. The proposed care support devices and systems can be divided into two groups: for outdoor and indoor use [5].

The systems for outdoor use usually employ devices with GPS receivers [6], which are used to localize the user outside his living place. Those systems usually employ a method known as geofencing, which allows to detect elopement behaviors such as leaving the designated areas or choosing a path, which is different than usual [7]. They can be also implemented using smartphones and detect basic types of activity [8].

The solutions intended for indoor use, due to lack of GPS signals inside buildings operate using different technologies ranging from ultra wideband [9] allowing for localization with errors of dozen centimeters to less accurate widely spread standards such as Wi-Fi [10], which allow to localize the users with accuracy of several meters. The data from those systems can be used to detect and assess wandering episodes. In the literature, there can be found different methods from that purpose ranging from simple room occupancy [11] to more detailed analysis of partitioned movement trajectories [12].

Wandering detection will be one of the functionalities of the platform being developed within the IONIS project [13]. It will be implemented using an indoor positioning system, deployed in elderly persons home. In this paper we present the results of initial field tests of Bluetooth Low Energy based localization system being a part of the IONIS platform. The main objective of the tests was to see, whether the system offers enough accuracy and reliability to properly diagnose and later, in the final product, detect wandering. The described tests were performed in a long term care center located in Warsaw.

The paper is structured as follows. Section II includes the description of the localization system and algorithm used during the experiments. Experiment results are presented in Section III. Section IV concludes the paper.

This work was supported by the National Centre for Research and Development, Poland under Grant AAL/Call2016/3/2017 (IONIS project).

Marcin Kolakowski is with the Institute of Radioelectronics and Multimedia Technology, Warsaw University of Technology, Poland (e-mail: m.kolakowski@ire.pw.edu.pl).

Bartosz Blachucki is with the Department of Geriatrics, Medical University of Warsaw, Poland (e-mail: bartosz.blachucki@wum.edu.pl).

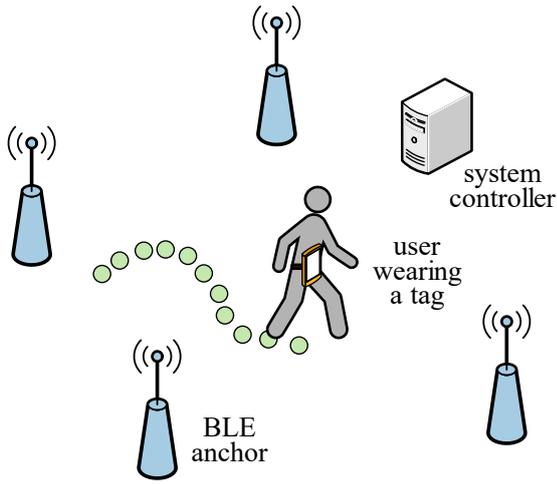

Fig. 1. Localization system's architecture

## II. LOCALIZATION SYSTEM

### A. System Architecture

The functional architecture of the localization system, which was used in the tests is presented in Fig. 1.

The system consists of three parts: localized device - tag worn by the user, an infrastructure comprising a set of anchors distributed in monitored area and a system controller.

The tag is a small, light device, which can be worn on a strap or in a pocket. It is equipped with a BL652 [14] Bluetooth 5.0 compliant radio module manufactured by Laird and is programmed to periodically broadcast packets in BLE advertisement channels. The packets are programmed to be transmitted at a constant rate of ten times per second.

The anchors receive the transmitted packets using two BL652 modules [14], which are equipped with external antennas of different polarizations placed a few centimeters apart from each other. Such solution allows to partially mitigate the impact of multipath propagation on measurement results. The anchors measure the signal strength of the received signals (RSS) and transmit the results to the system controller using Wi-Fi.

The system controller is a dedicated application running on a regular PC and is used to process the incoming measurement data. The main controller's functions are processing and storing measurement results, calculating user localization and providing the users with basic information on system's state. Besides that, it offers an interface, which allows for real-time visualization of the users locations.

### B. Localization Algorithm

The user position is calculated using an algorithm which employs the Extended Kalman Filter (EKF). The algorithm works in two phases consisting in predicting the users location based on his previous position and correcting the obtained value based on the measurement results supplied by the system. In the algorithm the user is modeled with a state vector, which contains information on his location coordinates and velocity:

$$x_k = [x \ v_x \ y \ v_y] \quad (1)$$

The vector includes users $x$, $y$ coordinates and velocity components $v_x$, $v_y$ from which movement direction and velocity can be estimated.

In the first phase of the implemented EKF the user's current location is predicted based on his previous coordinates and velocity. For prediction a Discrete White Noise Acceleration movement model [15] is used. In that model it is assumed that between the analyzed moments, the localized person moves along straight lines at a constant speed. The acceleration is treated as process noise. The prediction phase is implemented with the following formulas:

$$\hat{x}_{k(-)} = F\hat{x}_{k-1(+)} \quad (2)$$

$$P_{k(-)} = FP_{k-1(+)}F^T + Q_k \quad (3)$$

where $\hat{x}_{k(-)}$ is state vector prediction, $\hat{x}_{k-1(+)}$ state vector value in the previous iteration $P_{k(-)}$ and $P_{k-1(+)}$ corresponding covariation matrices, $F$ a matrix containing equations of motion and $Q_k$ is process noise matrix with values defined according to the DWNA model [15].

The obtained prediction is updated in the second phase using the measurement results obtained by the system infrastructure. Before feeding them to the algorithm, they are preprocessed by calculating the average received signal strength for each of the anchors. It is done in two steps. Firstly, for all of the received packets a mean result for both of the anchor BLE receivers is calculated. Then, the collected RSS measurement results are averaged per second. Such approach allows to mitigate the negative effects of multipath propagation, and in case of slowly moving elderly persons would not negatively impact localization accuracy. The averaged RSS results are then used to correct the predicted user location using the following dependencies:

$$z_k = [\ RSS_1 \ \cdots \ RSS_n\ ] \quad (4)$$

$$h_k(x_k) = [\ RSS_1(x_k) \ \cdots \ RSS_n(x_k)] \quad (5)$$

$$K_k = P_{k(-)}H_k^T\left(H_k P_{k(-)} H_k^T + R_k\right)^{-1} \quad (6)$$

$$\hat{x}_{k(+)} = \hat{x}_{k(-)} + K_k\left(z_k - h_k(\hat{x}_{k(-)})\right) \quad (7)$$

$$P_{k(+)} = (I - K_k H_k^T)P_{k(-)} \quad (8)$$

where $\hat{x}_{k(+)}$ is the final localization result, $z_k$ is a measurement vector containing the averaged RSS measurement results, $h_k(x_k)$ is the sensor model function which allows to estimate expected measurement results for the predicted user localization and $H_k$ is its linearized form, $K_k$ the Kalman gain and $R_k$ measurement covariance matrix.

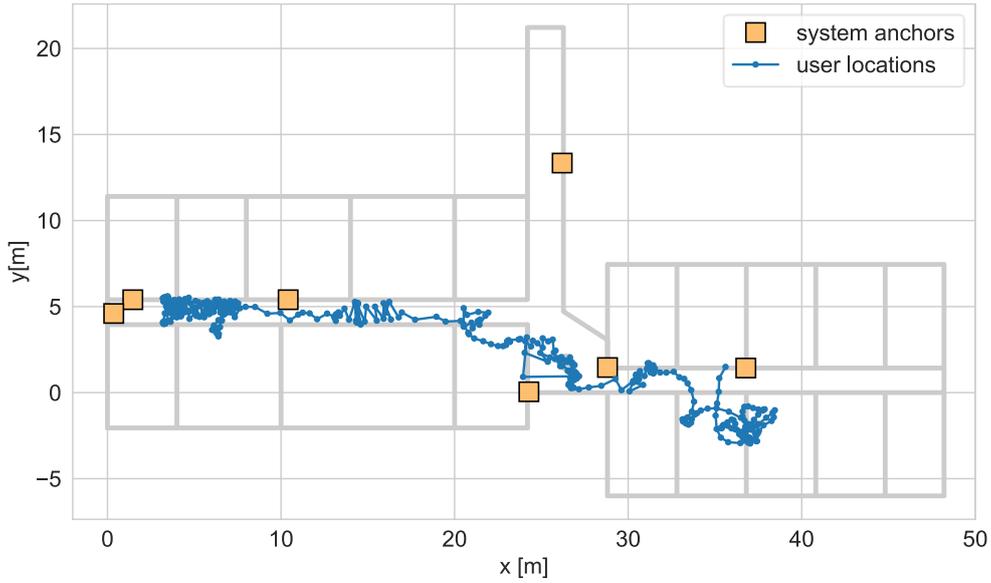

Fig.2 Experiment site and localization system layout with an exemplary registered trajectory

In the assumed sensor model, the power of the BLE signals received by the anchor $n$ is calculated using exponential path loss model:

$$RSS_n(x_k) = P_0 - 10\gamma \log_{10} \frac{d_n(x_k)}{d_0} \quad (9)$$

where $P_0$ is the RSS received at the reference distance $d_0$ (1 meter), $d_n(x_k)$ is the distance between the anchor $n$ and localization $x_k$ and $\gamma$ is path loss exponent (assumed 3.5).

The proposed algorithm allows for localization in two dimensions. It can be extended to handle three dimensional cases by adding $z$ coordinate and $v_z$ velocity component to the state vector. Such a solution however, due to moderate accuracy of BLE power measurements, might not give accurate results. In case, where the user elevation should be estimated, it would be better to implement floor detection rather than try to estimate his/her exact distance from the ground.

## III. EXPERIMENTS

The experiments were performed at one of health care facilities located in Warsaw. In the tests, four elderly people diagnosed with dementia were tested. All of them have manifested problems with wandering over their stay in the facility. That gave us a chance to see, whether the employed BLE based localization system could be used to properly observe and monitor this dementia symptom.

The layout of the facility, where the system was installed and the system infrastructure layout are presented in Fig. 2.

The test facility consisted of two long corridors and adjacent rooms. The patients did not often leave the floor, where their rooms were located, so the system was installed on one floor only. The infrastructure used in the experiment consisted of 7 anchors fixed to the walls at different heights (from 0.5 to 2.5 m). In order not to cause the other patients any inconvenience in their private spaces, the anchors were installed only in the corridors. Places, where the signal propagation conditions were more demanding were covered with multiple anchors.

During the tests, the patients were continuously tracked for 3 days. An exemplary trajectory registered using the localization system is presented in Fig.2. The system allows to localize the user with moderate accuracy – the trajectory passes through walls but it is enough accurate to give the general idea, where and how fast the user was moving.

During the experiments a few wandering incidents were registered. The exemplary registered wandering trajectories are presented in Fig. 3 and Fig.4.

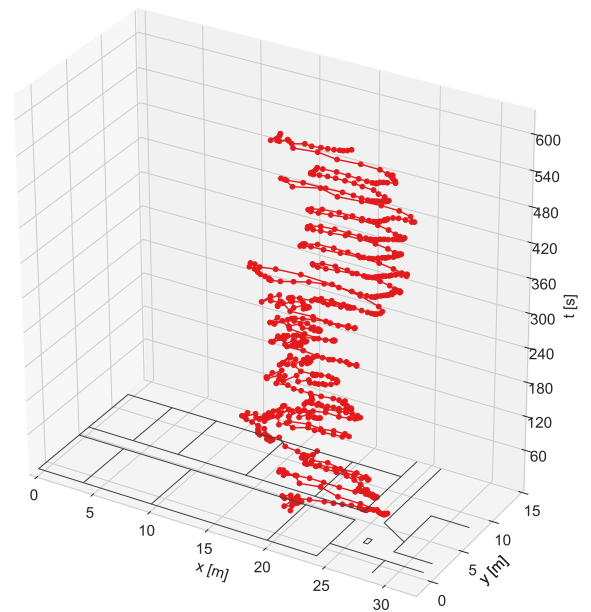

Fig. 3. Wandering trajectory 1 (400 m covered in 10 minutes)

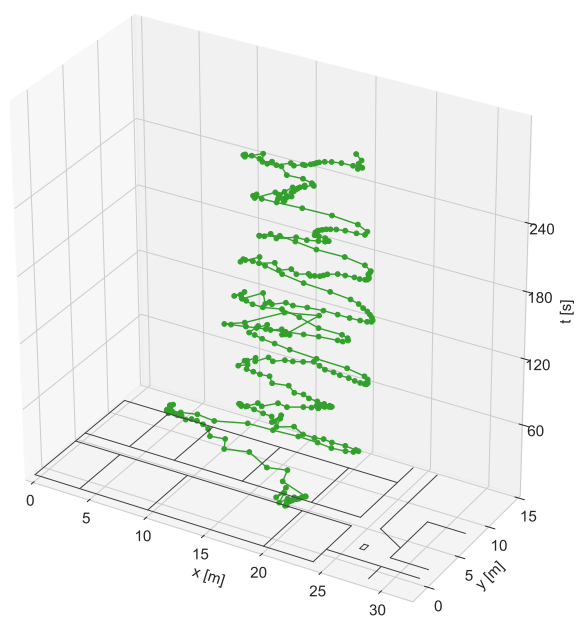

Fig. 4. Wandering trajectory 2 (250 m covered in 5 minutes)

Both of the presented wandering episodes happened in the corridors. In case of the first incident (Fig.3) the monitored patient repeatedly moved along a path of an approximate length of 7 meters. The incident took about 10 minutes and the patient covered a total distance of about 400 m.

The second detected wandering (Fig.4) occurred along a similar, but slightly longer (about 11 m) path. It took about 5 minutes, in which the patient covered 250 m.

The presented results prove that BLE based localization system developed within the IONIS project provides user localization, which is accurate enough to detect and diagnose wandering.

The presented results are the outcome of the initial field tests of the localization system developed within the IONIS project. The wandering incidents were recognized through analysis of the registered movement trajectories. In its final form, the IONIS platform will allow for automatic wandering detection.

## IV. CONCLUSIONS

In the paper the results of the experimental study concerning monitoring of wandering behaviors of people with dementia were presented. In the test a BLE based localization was used. The algorithm employed in the system was an implementation of the Extended Kalman Filter with additional mitigation of multipath effects.

The experiments were conducted in a long term care facility for four people with dementia. During the tests a few wandering episodes were registered. The localizations obtained with the system were accurate enough to observe movement trajectory repetitiveness, which is peculiar to wandering symptoms. It shows that the localizations calculated by the system might be useful for automatic wandering detection.

The presented results will be used in the future research tasks consisting in designing and developing methods for real-time wandering detection for use in the final IONIS platform.

ACKNOWLEDGMENT

We are very grateful and would like to offer our thanks for the opportunity to conduct the research in the long term care facility led by Congregation of Sisters Servants of the Immaculate Conception of the Blessed Virgin Mary.